# Engineering Hybrid Resonances in Nanophotonics

Shutao Zhang[1,2,3,#], Cheng-Feng Pan[1,4,#], Yandong Fan[1], Jehyeon Shin[6], Yuanda Liu[7], Yan Liu[2], Jun Ding[3], Jing Wu[8], Junsuk Rho[9,10,11,12], Yuri Kivshar[13,*], Joel K. W. Yang[1,5,*] and Zhaogang Dong[1,2,*]

[1]Singapore University of Technology and Design (SUTD), 8 Somapah Road, 487372, Singapore

[2]Quantum Innovation Centre (Q. InC), Agency for Science Technology and Research (A*STAR), 2 Fusionopolis Way, Innovis #08-03, Singapore 138634, Republic of Singapore

[3]Department of Materials Science and Engineering, National University of Singapore, 9 Engineering Drive 1, Singapore 117575, Singapore

[4]Department of Electrical and Computer Engineering, National University of Singapore, 4 Engineering Drive 3, Singapore 117576, Singapore.

[5]Singapore-HUJ Alliance for Research and Enterprise (SHARE), The Smart Grippers for Soft Robotics (SGSR) Programme, Campus for Research Excellence and Technological Enterprise (CREATE), Singapore 138602

[6]Graduate School of Artificial Intelligence, Pohang University of Science and Technology (POSTECH), Pohang 37673, Republic of Korea

[7]Institute of Materials Research and Engineering (IMRE), Agency for Science, Technology and Research (A*STAR), 2 Fusionopolis Way, Innovis #08-03, Singapore 138634, Republic of Singapore

[8]School of Electronic Science and Engineering, Southeast University, Nanjing, 211189 China

[9]Department of Mechanical Engineering, Pohang University of Science and Technology (POSTECH), Pohang 37673, Republic of Korea

[10]Department of Chemical Engineering, Pohang University of Science and Technology (POSTECH), Pohang 37673, Republic of Korea

[11]Department of Electrical Engineering, Pohang University of Science and Technology (POSTECH), Pohang 37673, Republic of Korea

[12]POSCO−POSTECH-RIST Convergence Research Center for Flat Optics and Metaphotonics, Pohang 37673, Republic of Korea

[13]Research School of Physics, Australian National University, Canberra ACT 2601, Australia.

[#]These authors equally contribute to this work.

*Correspondence and requests for materials should be addressed to Z.D. (email: zhaogang_dong@sutd.edu.sg), J.K.W.Y. (email: joel_yang@sutd.edu.sg), and Y.K. (email: yuri.kivshar@anu.edu.au).

**Abstract**

Hybridization of resonances is known to overcome inherent limitations of individual systems, enabling advanced functionalities and applications. Here we discuss hybrid plasmonic-Mie resonators that emerged recently as a promising direction in advancing nanophotonic structures by synergistically combining the strong near-field enhancement of plasmonic components with the low-loss, multipolar resonances of dielectric Mie elements. We review the recent progress in the field, encompassing the fundamental physical principles, structural design strategies, material platforms, computational optimization approaches, and representative device implementations. Our discussion starts by evaluating the complementary characteristics of plasmonic and Mie resonances followed by a description of the coupling between these resonances in order to boost light-matter interactions. Afterward, we explore the performance of efficient hybrid resonators for different application areas. Apart from the conventional metal-dielectric systems, we consider the recent class of epsilon-near-zero (ENZ) materials, which can provide unique advantages in terms of field localization, phase engineering, and energy flow management in the vicinity of zero-permittivity conditions, offering more flexibility in designing hybrid nano-optical devices. Lastly, we point out potential research avenues aiming to improve functional and efficient nanophotonic devices, especially those involving emerging topological material systems, such as $Sb_2Te_3$, $Bi_2Te_3$, $Bi_2Se_3$, combining plasmonic amplification, dielectric confinement, and spin-dependent optical behavior.

## 1. Introduction

In recent decades, the nanophotonics paradigm has proven to be a powerful tool for controlling the interaction between light and matter at subwavelength scales, leading to the development of ultra-small, highly integrated, and multifunctional optical devices.[1-4] This technology is being applied in various fields, from optical displays[5-11] to photodetectors[12-16] and sensors[17-19], through spectrometers,[20,21] optical imaging tools, nonlinear optics,[22-24] waveguides,[25] and even quantum photonics.[26-30]

Optical resonances, in turn, can play a key role in many of the above applications due to their capability of confining, amplifying, and controlling electromagnetic waves at the nanoscale.[31-33] For example, plasmonic resonances take advantage of free electron oscillations occurring in metallic nanostructures, allowing for deep sub-wavelength confined and accompanying strongly enhanced near-fields.[34-36] Conversely, high index dielectric resonances are based on geometry-dependent Mie-type multipoles with low optical losses.[37,38] Nonetheless, while presenting attractive features, plasmonics and high-index dielectrics stand as complementary but not ideal alternatives. On one hand, the highly confining nature of plasmonic resonance results in strong intrinsic dissipation, especially at the visible and near-infrared spectra. On the other hand, high-index dielectric resonators offer lower optical loss but suffer from lower hotspot intensity and confinement.[39] Other related areas include the study of epsilon-near-zero (ENZ) materials that allow for novel permittivity distributions that can affect the flow of optical energy.

Hybridizing both plasmonic and dielectric modes has been a promising approach to overcome some of these limitations, combining subwavelength confinement, minimized energy losses, and rich modal composition within a single structure.[40-42] Such a strategy can be described by the following three aspects: materials and resonant platform consisting of plasmonic materials, dielectrics, ENZ materials, and hybrid materials; hybrid mode-level phenomena, including broad plasmonic responses combined with geometrically defined dielectric multipole resonances through interference and loss redistribution; and optical functions provided by near-field enhancement, radiation control, light-matter interaction, or nonlinear/frequency-conversion emission. Therefore, hybridizing plasmonic and dielectric modes allows for the creation of not only a single device but a nanophotonic framework where confinement and linewidth, modal composition, and functionality become designable factors.

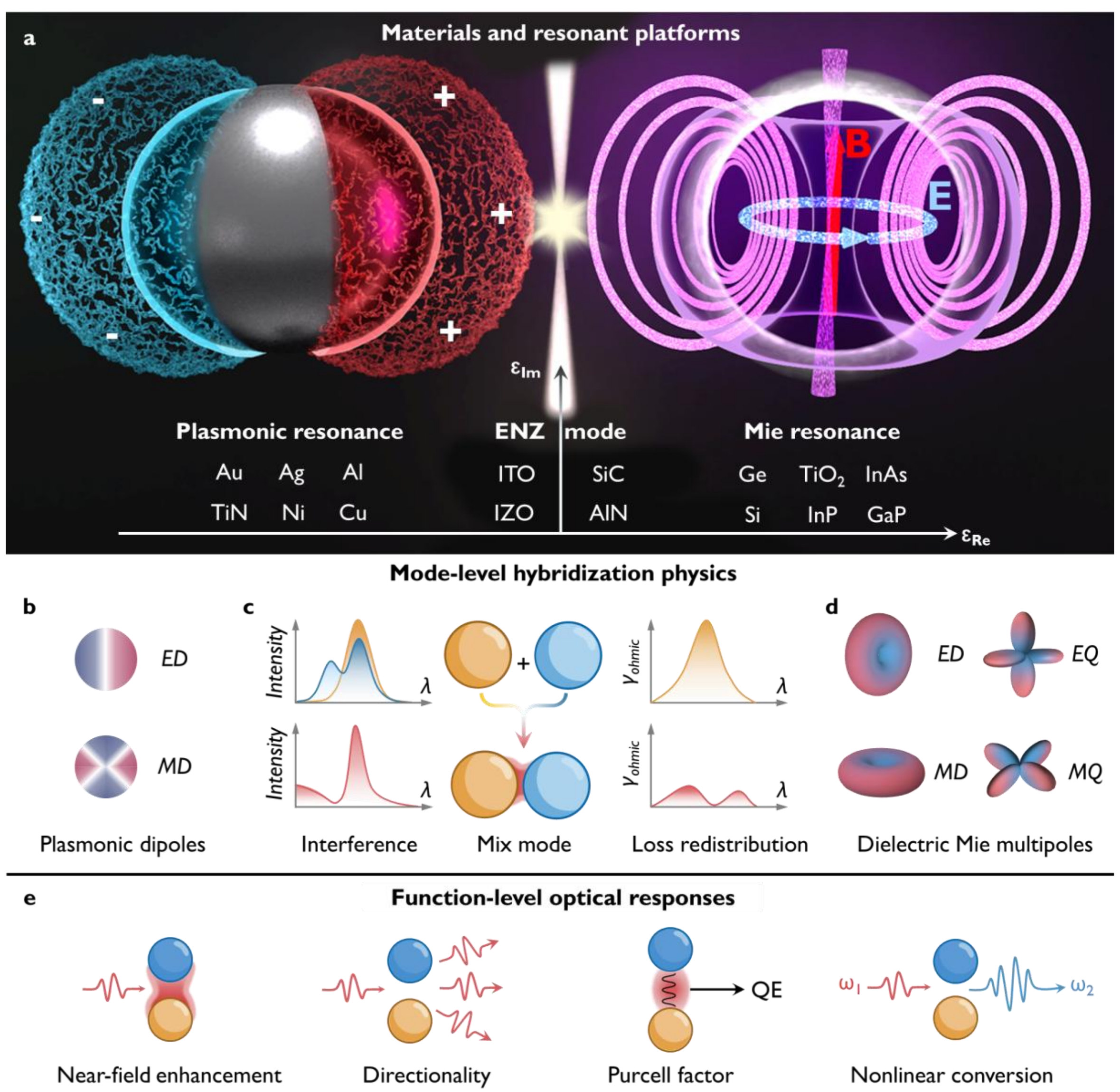

**Figure 1. Hybrid plasmonic–Mie resonators: Material platforms, modal framework, and function-level optical responses. a.** Material landscape mapped by the real part of the permittivity ($\varepsilon_{\mathrm{Re}}$), spanning plasmonic materials ($\varepsilon_{Re} < 0$; for example, Au, Ag, Al, and TiN), epsilon-near-zero (ENZ) media ($\varepsilon_{Re} \approx 0$; for example, ITO, IZO, SiC, and AlN), and dielectric platforms supporting Mie resonances ($\varepsilon_{Re} > 0$; for example, Si, Ge, $TiO_2$, and GaP). **b.** Plasmonic modal responses, represented by broad electric- and magnetic-dipole resonances. **c.** Hybrid plasmonic–Mie modes arising from modal coupling, characterized by interference-enhanced resonances and redistribution of loss pathways. **d.** Dielectric Mie modes with multipolar character, including ED, MD, EQ, and MQ responses. **e.** Function-level optical responses accessible in these systems, including near-field enhancement, radiation shaping, engineered light–matter interaction, and nonlinear or frequency-converted emission.

## 2. Mechanisms of Hybridization

Hybrid plasmonic-Mie resonances result from the intentional coupling of complementary modal classes, especially the classically loss-free Mie resonances in dielectrics and the strongly localized plasmonic resonances, where ENZ modes bring yet another dispersive dimension in extended hybrids. Hybrid resonances go beyond simply combining metals with dielectrics into the same nanostructure, but to a broader approach wherein all three dimensions of linewidth, field localization, and phase can be combined into one resonator structure. Figure 2 illustrates this concept, progressing from elementary modal pieces to hybrid linewidth control and, finally, phase-engineered multifunctionality.

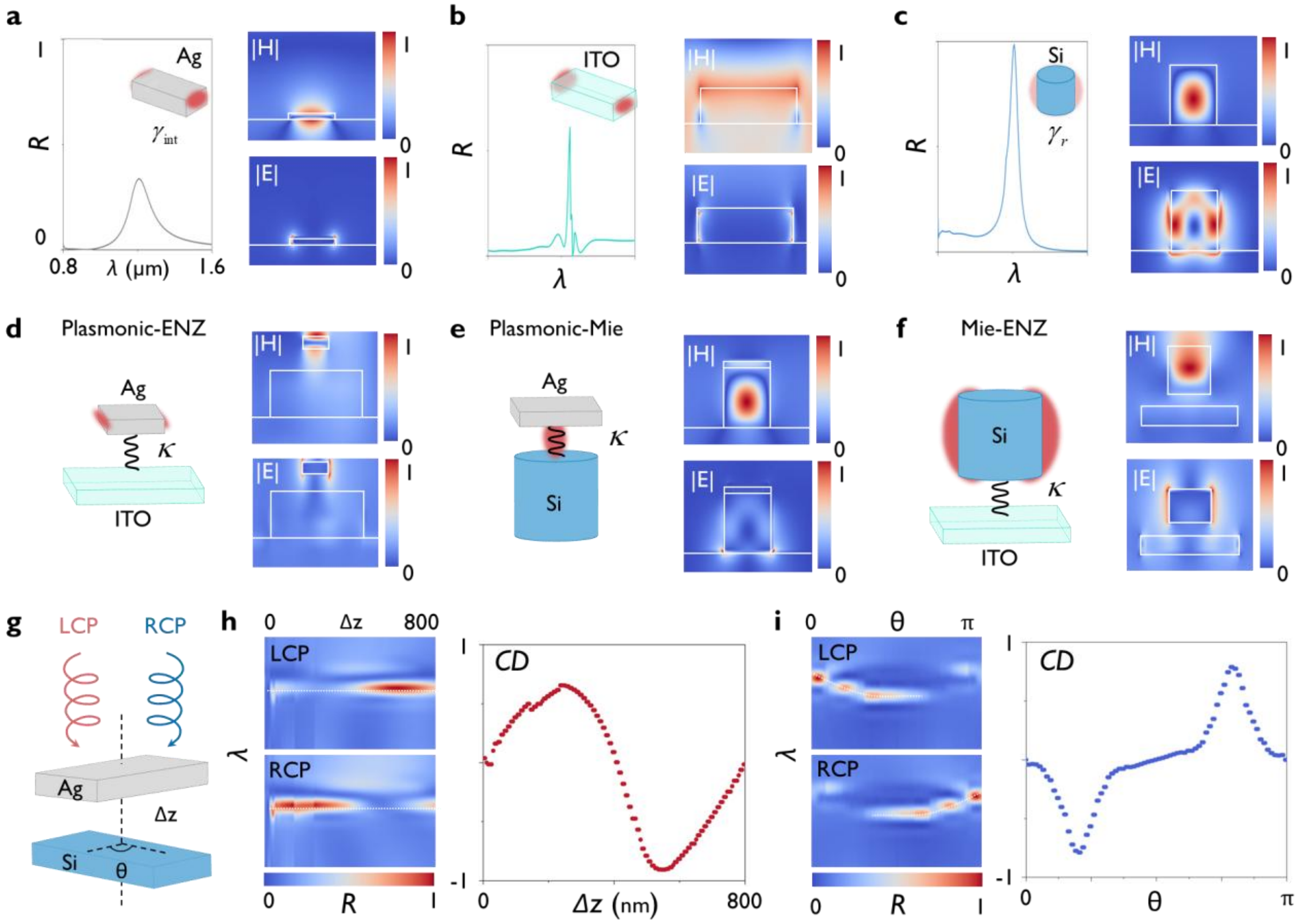

**Figure 2. Material-specific resonances, hybrid-mode formation, and chiral validation. a-c.** Fundamental building blocks: **a.** Plasmonic Ag (high confinement, intrinsic loss $\gamma_{int}$), **b.** ITO ENZ (epsilon-near-zero response), and **c.** Dielectric Si (low-loss Mie resonance, radiative loss $\gamma_r$). **d-f.** Hybridized coupling regimes: Near-field maps (|*H*|, |*E*|) illustrating three hybrid pathways: **d.** Plasmonic-ENZ: Interaction between localized plasmons and ENZ modes. **e.** Plasmonic-Mie: Hybridization of dissipative metallic and high-Q dielectric modes to balance field enhancement

and coherence. **f.** Mie-ENZ: Coupling between dielectric Mie dipoles and the ENZ environment. **g.** Geometric phase engineering: A hybrid Ag-Si meta-atom where rotation angle $\theta$ and separation $\Delta z$ tune the relative phase between coupled modes.**h, i.** Chiral response and mechanism: Circular dichroism (CD) spectra showing sinusoidal dependence on $\theta$ and $\Delta z$. Field distributions: Near-field interference under LCP and RCP excitation, confirming the phase-engineered origin of the chiral response.

### 2.1. Complementary modal building blocks

Dielectric and plasmonic resonators represent two limiting regimes of nanoscale light confinement, as shown in Fig2.a-c. In high-index dielectric nanoparticles, optical scattering can be understood in terms of electric and magnetic multipoles, whose collective contribution may be written as [43,44]

$$C_{sca} \propto \sum_{l=1}^{\infty}(2l+1)(|a_l|^2+|b_l|^2) \tag{1}$$

where $a_l$and $b_l$denote electric and magnetic multipolar coefficients, respectively. In contrast to metallic nanostructures, dielectric resonators can efficiently support both electric and magnetic modes with relatively low dissipative loss, which generally leads to narrower linewidths and higher quality factors.[39,43,45] However, the associated optical energy is often distributed over a larger mode volume, limiting the strength of local field concentration.

Plasmonic resonators occupy the opposite regime. Their optical response is dominated by collective oscillations of free electrons, which compress electromagnetic energy into deeply subwavelength volumes and generate intense local hotspots.[46] Although such confinement can be extremely strong, it usually comes at the cost of pronounced intrinsic dissipation, which leads to resonance broadening and a loss of spectral coherence. The ENZ modes represent an entirely different type of dispersive regime, where the near-zero value of the permittivity affects field compression and phase development, and neither can be explained solely by either metallic or dielectric phenomena. In such a case, dielectric and plasmonic modes become a complement for each other, wherein dielectric resonances offer spectral selectivity and plasmonic resonances enable maximum field localization. The combination of plasmonic and Mie resonances into hybrid plasmonic-Mie resonance is achieved upon careful matching of both frequencies and mode distributions, as illustrated in Fig. 2d-f.

## 2.2. **Hybridization-enabled redistribution of linewidth and confinement**

The essential physics of hybrid plasmonic-Mie resonances can be captured by a minimal coupled-mode description involving a dielectric mode $(\omega_p, \gamma_r)$ and a plasmonic mode $(\omega_p, \gamma_{int})$,[47]

$$\begin{pmatrix} \omega_d - i\gamma_r/2 & K \\ K & \omega_p - i\gamma_{int}/2 \end{pmatrix} \begin{pmatrix} a_d \\ a_p \end{pmatrix} = 0 \tag{2}$$

The value of $K$ represents the coupling strength between the two modes, $\gamma_r$ is the radiative linewidth of the dielectric resonance, and $\gamma_{int}$ is the intrinsic linewidth of the plasmonic resonance. The magnitude of $K$is set by the modal overlap between the uncoupled modes, requiring spectral proximity, near-field co-localization, non-orthogonal polarization components, and compatible symmetry.[48,49] In this context, the hybrid eigenmodes can be expressed as follows

$$\widetilde{\omega}_\pm = \frac{\widetilde{\omega}_d + \widetilde{\omega}_p}{2} \pm \sqrt{K^2 + \left(\frac{\widetilde{\omega}_d - \widetilde{\omega}_p}{2}\right)^2} \tag{3}$$

This hybridization must not be seen as an overlay of two separate resonances. On the contrary, it is the way in which the optical energy concentration as well as the linewidth division can be redone inside the hybrid structure. As far as the dielectric regime goes, the dominant source of losses here will still be radiative leakage, leading to high Q-factor values. In the plasmonic limit, the field is compressed much more strongly, but at the expense of increased intrinsic dissipation and a broader linewidth. In the hybrid regime, the effective linewidth can be approximately viewed as a weighted combination of the two loss channels,

$$\Gamma_\pm \sim \alpha_\pm \gamma_r + \beta_\pm \gamma_{int} \tag{44}$$

where $\alpha_\pm$and $\beta_\pm$depend on the modal composition of the hybridized states. The resulting mode inherits radiative coherence from the dielectric constituent and deep subwavelength confinement from the plasmonic constituent, enabling a tunable balance between quality factor and local enhancement.[50] In extended hybrid systems, ENZ-assisted coupling can further perturb this balance by modifying the local permittivity environment and modal energy flow. As schematically illustrated in Fig. 2d,f, the significance of hybridization lies not in eliminating the trade-off between linewidth and confinement, but in rendering this trade-off engineerable through spectral detuning, geometry and coupling strength.

### 2.3. **Hybridization-enabled redistribution of linewidth and confinement**

Beyond linewidth and confinement control, hybrid plasmonic-Mie resonators possess a further degree of freedom: the relative phase between coupled dipolar responses. Once dielectric and plasmonic modes are hybridized within an anisotropic meta-atom, geometry can be used not only to tune the coupling strength, but also to control the phase relation between the participating dipoles. In the hybrid structure illustrated in Fig. 2g, the in-plane rotation angle $\theta$ and out-of-plane separation $\Delta z$ jointly determine the relative phase,

$$\Delta\phi = \theta - k_z \Delta z \quad (5)$$

where $\theta$ describes the orientation-induced phase contribution and $k_z \Delta z$ represents propagation-induced phase accumulation along the vertical direction. This formulation provides a compact physical picture linking geometric design to interference-driven optical functionality.

When the coupled dipoles exhibit comparable amplitudes and a relative phase approaching quadrature, the hybrid resonance becomes strongly chiral.[51] This immediately explains the sinusoidal dependence of circular dichroism on both $\Delta z$ and $\theta$, as shown in Fig. 2h,i. Both parameters modulate the same underlying phase difference. Hybrid plasmonic-Mie resonators should therefore be regarded not merely as linewidth-optimized cavities, but as multifunctional coupled-dipole platforms in which loss channels, field localization and phase-sensitive optical responses can be co-engineered within a single nanophotonic system.[52]

Taken together, this picture suggests that hybrid plasmonic–Mie resonances provide a unified framework for nanophotonic design, where modal complementarity, linewidth and loss redistribution, and phase control can all be realized within a single meta-atom. This framework provides the physical basis for the application domains discussed in the following sections: spectral and computational design, energy conversion and carrier dynamics, radiative and directional emission, and emerging material platforms. Representative hybrid systems are summarized in Table 1, together with their material combinations, coupling mechanisms and functional advantages. These examples underscore the versatility of hybrid plasmonic-Mie coupling as a general design strategy for multifunctional nanophotonic devices.

**Table 1. Representative hybrid plasmonic-Mie material systems, physical mechanisms, and applications.**

| Application | Structure / Materials | Mechanism | Advantages | Ref. |
| --- | --- | --- | --- | --- |
| Display / Coloration | Si-Al, Si-Cr nanodisks | Metal-enhanced Mie *MD* | High color purity | [53,54] |
| | Ag-Si metasurface | Electrical plasmon-Mie coupling | Tunability | [55] |
| | Au-$TiO_2$ / Si-Au gratings | Diffractive-plasmon-Mie hybridization | Multi-state coloration | [56,57] |
| Photodetection / Sensing | *p*-Si/Al Schottky hybrid | Mie-plasmon + Schottky extraction | Spectral / polarization selectivity | [58] |
| | Si-Au nanocavity | Metal-dielectric-metal absorption | Broadband absorption | [59] |
| | $Sb_2Te_3$ metasurface (TI) | Topological plasmon-Mie coupling | VIS-MIR broadband response | [60] |
| | Si-Ag microring / slot | Hybrid guided-mode resonance | High RI sensitivity | [61,62] |
| Photoemission / Modulation | Si NP-Au mirror | Image-dipole interference | Purcell enhancement | [63,64] |
| | Au-Si, GaAs Yagi-Uda | Plasmon-Mie dipole interference | Directionality / polarization control | [65,66] |
| | $Sb_2Te_3$-Perovskite disk | Hot-electron injection | Electrically tunable emission | [67] |
| On-Chip Photonics | Ge-metal cavity, Si-Ag microring | Hybrid cavity mode | High-Q / integration | [61,62,68] |
| | α-Si:H-Au metasurface | Coupled Mie-SPP resonance | Polarization control / switching | [69,70] |
| Energy / Magneto-Optics | $TiO_2$-Au-CdS composite | Mie-assisted plasmonic photocatalysis | Enhanced light-harvesting / $H_2$ generation | [71,72] |

| Application | Structure / Materials | Mechanism | Advantages | Ref. |
| --- | --- | --- | --- | --- |
| | Si-Fe-dielectric stack | Magnetic Mie resonance | Magneto-optic modulation | 73 |

### 3. Computational Design and Spectral Engineering of Hybrid Resonators

Machine learning (ML) has become a powerful framework for accelerating the design and analysis of hybrid nanophotonic resonators by rapidly predicting optical responses and extracting physically meaningful patterns from large simulation or experimental datasets.[74-77] Rather than replacing conventional electromagnetic solvers such as finite-difference time-domain (FDTD) and finite-element method (FEM), ML serves as a surrogate framework that shortens the design cycle by screening candidate structures, approximating optical responses, and guiding optimization. In high-dimensional spaces involving multiple geometric and material parameters, ML surrogates can map resonator configurations to spectra, field distributions, and modal descriptors with orders-of-magnitude speedup over full-wave solvers while retaining high predictive accuracy. [78]

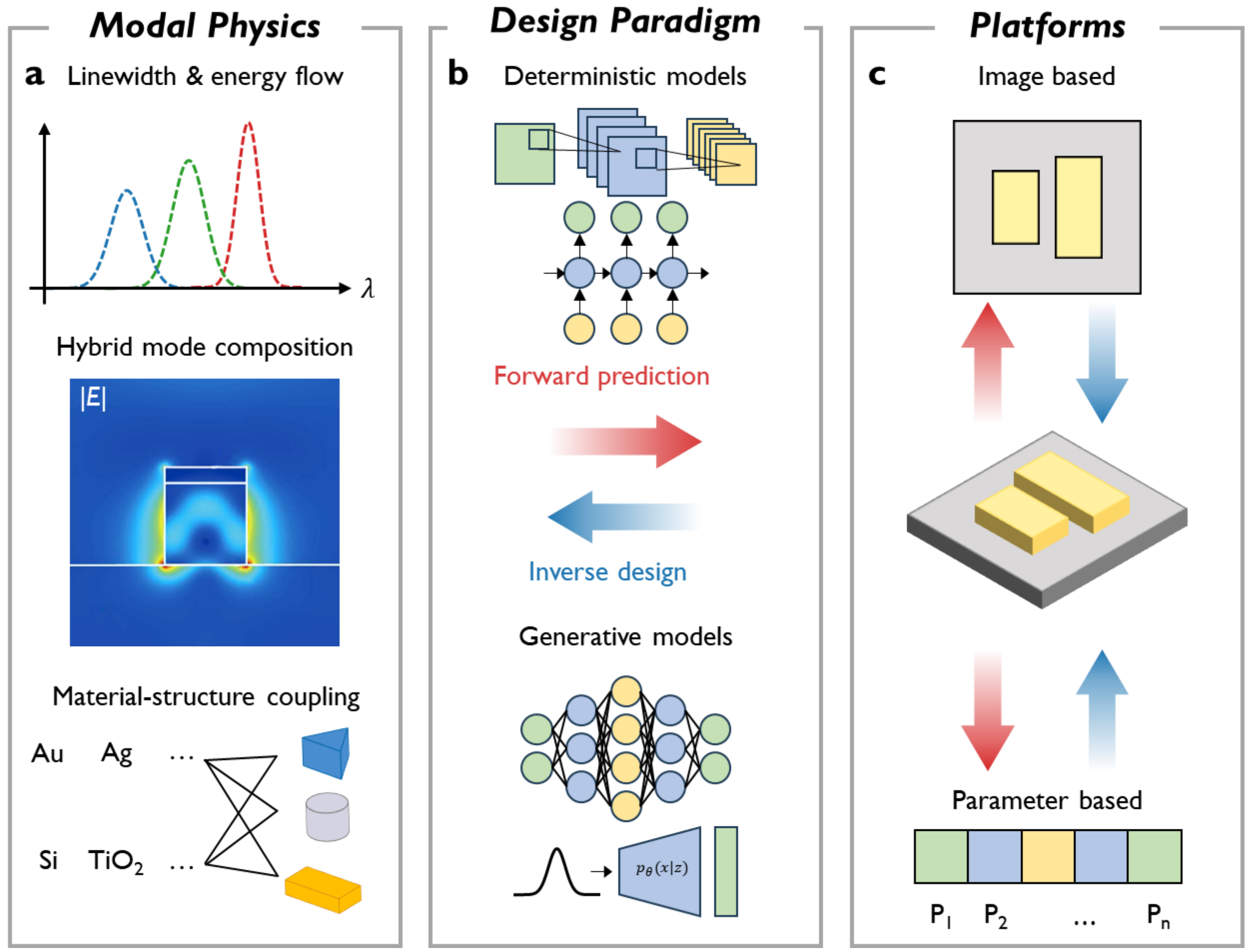

**Figure 3. Machine-learning framework for hybrid plasmonic-Mie nanophotonics.** Machine learning contributes to hybrid plasmonic-Mie nanophotonics across four interconnected aspects: a**.** modal physics, including linewidth and energy flow, hybrid mode composition, and material-structure coupling; **b.** design paradigm, encompassing deterministic and generative models for forward prediction and inverse design; and **c.** platforms, where metasurfaces can be represented in image-based or parameter-based forms.

### 3.1. Functional Targets

From the functional perspective, ML provides an efficient route to predict and optimize device-level optical responses relevant to filtering, sensing, emission control, and energy conversion. Supervised models can map structural parameters to absorption, transmission, reflection, and scattering spectra, enabling rapid spectrum-level evaluation without repeatedly solving the full electromagnetic fields.[76,79,80] This is particularly valuable for broadband absorbers, wavelength-selective resonators, and high-Q spectral devices, where brute-force parameter sweeps are prohibitively expensive.[80] ML can also target higher-level functionalities, including near-field enhancement, radiation engineering, light-matter interaction, and nonlinear photonics. It therefore shifts hybrid nanophotonic design from trial-and-error optimization to function-driven search.

### 3.2. Learning Hybrid Modal Physics

Beyond functional prediction, ML is increasingly being used to learn the modal physics of hybrid nanophotonic systems. Spectra provide a compact measure of performance, but field distributions and modal decompositions reveal how plasmonic and Mie resonances hybridize, where energy is localized, and how interference governs absorption, emission, and scattering.[80] As summarized in Fig. 3a, ML can capture descriptors such as linewidth and energy flow, hybrid mode composition, and material-structure coupling. Full-field prediction is particularly important for applications governed by local field distributions, including nonlinear optical processes controlled by field overlap,[79] Purcell enhancement at emitter placements,[81] and directive radiation engineering.[45,82] Because the output space is high-dimensional and resolution-dependent,[83] recent work has moved toward more structured learning strategies.[84-86] Physics-guided neural networks impose Maxwell-consistent constraints to improve data efficiency and physical consistency,[85] whereas operator-learning approaches aim to learn resolution-invariant mappings from geometry and material distributions to complex field patterns. [86]

### 3.3. Design Paradigms and Platform Representations

These capabilities directly enable new design paradigms for hybrid nanophotonics. In the forward direction, ML surrogates rapidly evaluate candidate structures; in the inverse direction, they identify geometries or material combinations that satisfy target spectral, field, or modal objectives. As shown in Fig. 3b, this framework includes both deterministic models and generative models for forward prediction and inverse design. A central challenge in inverse nanophotonic design is the intrinsic non-uniqueness of the mapping from target response to structure. Tandem neural-network schemes help alleviate the one-to-many nature of the inverse design problem by combining an inverse predictor with a forward surrogate model. In doing so, they improve the likelihood that the generated solutions are not only physically reasonable but also consistent with the intended optical response.[79,87] Generative models, including VAEs, GANs, and more recently diffusion models, push this strategy further by enabling the sampling of multiple candidate structures under the same target response.[88-90] These candidate designs can be evaluated rapidly using the forward model and subsequently refined with standard electromagnetic solvers. Such data-driven design is particularly attractive for hybrid resonators, where competing plasmonic and dielectric modes, multiple geometrical degrees of freedom, and fabrication constraints make intuition-driven optimization difficult. As indicated in Fig. 3c, these workflows can be implemented on metasurface platforms represented in image-based or parameter-based forms, which in turn shape model architecture, design flexibility, and fabrication relevance.

## 4. Applications for Nanophotonic Devices and Systems

Hybrid Plasmonic-Mie resonance devices have attracted significant attention in the fields of photonics and nanotechnology. These devices are composed of carefully engineered subwavelength structures, such as nanoantennas, nanowires, or nanoparticles-designed to manipulate and control light at the nanoscale.[91] Hybrid Plasmonic-Mie resonance devices employ co-designed subwavelength metal-dielectric elements (*e.g.*, nanoantennas with metallic caps or back reflectors, core-shell nanoparticles, gap-plasmon cavities, and metal gratings) to couple field-enhancing plasmonic modes to high-Q dielectric multipoles. A key advantage of hybrid plasmonic-Mie structures lies in their ability to combine Mie resonances with plasmonic enhancements, offering strong local field confinement, tunable spectral responses, and additional optical functionalities beyond those achievable with pure dielectric Mie resonators. This synergy has led to improved performance across a range of applications, such as advanced displays, photodetectors, photoemitters, and integrated photonic waveguides. In the following sections, we discuss how hybrid plasmonic–Mie resonances have been exploited in different classes of nanophotonic devices.

### 4.1. Spectral Engineering with Hybrid Nanophotonic Resonators

The tuning of the spectral response of light at the nanoscale is one of the main goals of nanophotonics, having applications to imaging, sensing, display devices, and optical signal processing. In addition, it provides an excellent platform for studying interactions between light and matter in complex materials. With respect to that, hybrid plasmonic-Mie (HPM) resonators form an intriguing approach. Hybridizing plasmonic components' ability to localize and couple with light and Mie resonators' spectral selectivity and low losses leads to an impressive level of flexibility in designing the desired spectral features such as resonance wavelength, linewidth, and modulation depth.

As illustrated in Fig. 4, HPM resonator arrays can selectively receive broadband incident light and redistribute it into designed spectral output channels. From this perspective, spectral engineering based on HPM systems can be discussed in three main categories: static spectral engineering, dynamic spectral control, and high-Q spectral engineering.

Among the different implementations, disk-type architectures are among the most widely studied platforms for static spectral engineering. In these structures, dielectric resonators are integrated with metallic elements so that spectrally selective Mie-like resonances are combined with plasmon-enhanced field confinement. This combination is especially useful for structural coloration, where densely packed resonators often experience unwanted near-field coupling and reduced spectral stability. A representative example is shown in Fig. 4b, where Nagasaki *et al.* introduced an additional Cr layer to suppress optical coupling between neighboring pixels in high-resolution hybrid color arrays, thereby stabilizing the spectral response and preserving color fidelity even at high packing densities.[92] Other hybrid disk- and pillar-based designs have likewise shown that metallic caps or back reflectors can improve spectral selectivity, narrow the resonance features, and maintain high reflectivity.[53, 54] These results establish HPM structures as an effective platform for static color generation.

Beyond these more conventional geometries, alternative configurations such as gratings and nanoparticle-based hybrid systems further broaden the design space by introducing diffractive effects or scattering-mediated coupling. In the grating-assisted structure shown in Fig. 4c, Kim *et al.* combined dielectric nanodisks, metallic layers, and periodic gratings to support Mie, plasmonic, and diffractive resonances simultaneously.[56] This multimode hybridization enabled multiple color states and multilevel optical information encoding within a single pixel array, highlighting the potential of HPM structures for static spectral manipulation beyond simple coloration. Related nanoparticle-based systems have also demonstrated tunable and angle-insensitive color generation through hybrid plasmonic-Mie scattering.[57]

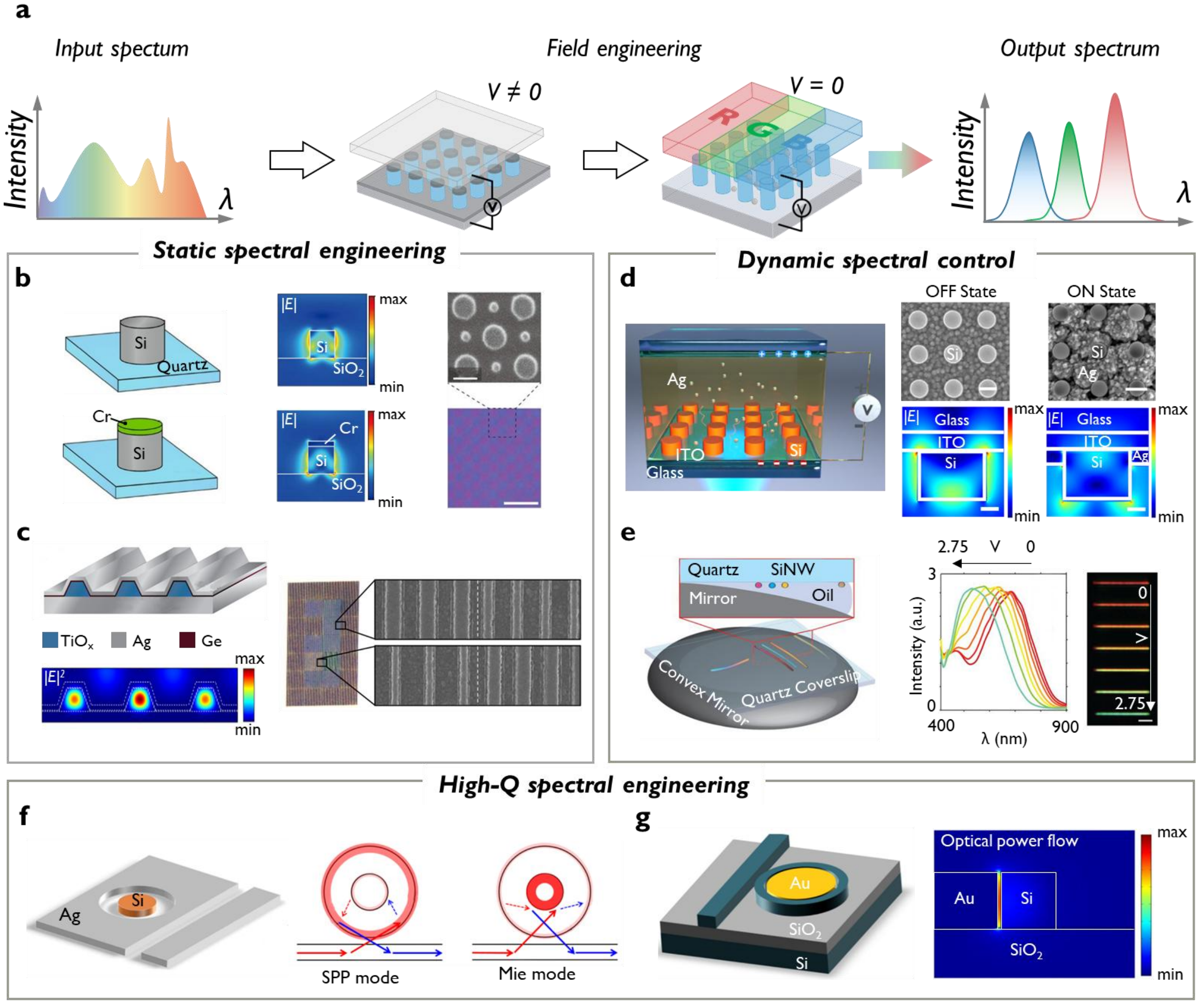

**Figure 4. Hybrid plasmonic-Mie cavities for nanostructural static and dynamic display technologies. a**. Schematic of HPM spectral engineering. A broadband input spectrum is incident on a hybrid plasmonic-Mie resonator array, where tailored hybrid resonances selectively collect and redistribute different wavelength components. **b**. The schematic of Si nanodisk and Cr-mask Si nanodisk. On the right is the field distribution of a single Mie resonator, and the SEM and microscope image of the periodic pattern is on the left.[92] **c**. A schematic illustration of a 1D nanoresonator grating featuring periodically spaced $TiO_x$ nanowires on an Ag substrate, encapsulated beneath a sub-1nm-thick Ge layer and a 30 nm-thick Ag film. On the right are the microscope and SEM images.[56] **e.** Schematic of the reversible electrical switching device, field distribution and SEM images.[55] **f.** Schematic of a nano-electromechanically (NEMS) tunable hybrid plasmonic-Mie resonator. A suspended silicon nanowire is positioned above an aluminum mirror, forming a tunable vertical cavity. By adjusting the bias voltage, the nanowire-mirror gap is modulated, allowing real-time control over the Mie-like resonance condition and enabling

continuous spectral tuning across the visible range.[93] **g.** The schematic of the Mie resonance nanofilter and the illustration of loss mechanism in traditional nanocavity and Mie resonance nanocavity.[68] **h.** Schematic structure of the proposed hybrid plasmonic waveguide-based microring sensor and average optical power flow in the propagation direction for $W_{slot}$ = 10 nm.[62]

In contrast to static designs, dynamic spectral engineering enables actively reconfigurable optical responses. This functionality often arises from the metallic component of the hybrid structure, which allows electrical, electrochemical, or mechanical tuning of the resonant condition. As shown in Fig. 4d, Zhang *et al.* demonstrated an electrically switchable HPM color platform through the reversible electrodeposition of silver nanoparticles on silicon nanostructures.[55] The reversible formation of the metallic nanoparticles dynamically perturbed the hybrid plasmonic–Mie resonance, enabling real-time switching between vivid primary colors and a nearly colorless state. A second tunability approach is presented in Fig. 4e where Holsteen et al. created a novel nanoelectromechanical system involving a suspended silicon nanowire placed above an aluminum mirror.[93] By adjusting the distance with nanometer accuracy under bias conditions, the authors were able to continuously change the interference conditions and the corresponding Mie-like resonance, resulting in significant redshift in the visible region. These examples demonstrate that HPM resonators may also allow both discrete and continuous spectral tuning, providing an additional capability in active photonic devices and displays.

High-Q spectral engineering can be additionally addressed using HPM resonators to create hybridization-induced narrowband and highly selective optical responses. The combination of plasmonic confinement and modal selectivity enables miniaturization of photonic components that act as filters, couplers, and sensors. As an example of the hybrid spectral engineering, the nanofilter presented in Fig. 4f was introduced by Yao et al. who incorporated a high-index germanium disk inside a plasmonic nanocavity achieving a very high Q-factor of up to 1292 and simultaneously spectral tunability in the telecommunication wavelength range.[68] Due to plasmonic effects, the system provided high field confinement, whereas the dielectric disk retained resonant modal selectivity, enabling linewidth tuning from 14.2 nm to 1.2 nm. This example captures the main strength of the hybrid architecture: combining small volume with sharp spectral peaks in one structure. Similar results can be seen in hybrid ring/microring resonators shown in Fig. 4g. It was demonstrated that the plasmon-enhanced whispering gallery/gap modes in such resonators may provide high refractive index sensitivity in miniature devices.[61, 62]

In summary, spectral engineering using hybrid plasmonic-Mie resonators has proven itself as an effective approach including static spectral shaping, spectral modulation, and high-Q spectral control. Hybrid spectral engineering utilizes the unusual opportunities that arise due to interaction of different optical modes and allows controlling the spectral response while keeping compact form factors. Future progress in HPM resonator development will be likely driven by advances in materials, fabrication technology, and design techniques.

### 4.2. Energy Conversion and Carrier Dynamics in Hybrid Nanophotonic Systems

When light is selectively confined within nano-optical structures, the absorption of energy no longer remains solely in the optical process. Instead, it will be further transformed and distributed to other processes such as electron excitation, carrier generation, transport and relaxation, ultimately manifesting as readable characteristic functional responses at the device level. In this sense, the hybrid nano-optical system provides an intrinsic bridge for the extension of spectral control to the realization of optical physical functions. As illustrated in Fig. 5a, incident light is first harvested through resonant absorption, followed by energy conversion into nonequilibrium states such as localized heating, electronic excitation, or hot-carrier generation. These processes then evolve into carrier separation, transport, and relaxation, leading to measurable signal generation. On this basis, this section discusses four representative directions: spectral-selective photodetection (Fig. 5b,c), geometry-driven absorption engineering (Fig. 5d,e), interference-assisted absorption control (Fig. 5f,g), and hot-carrier-mediated energy conversion (Fig. 5h-j).

Spectral-selective photodetection exploits the spectral separation of plasmonic and Mie resonances within a single hybrid device to achieve wavelength-dependent functionality. In the dual-mode detector shown in Fig. 5b, Zhang et al. designed a hybrid $Sb_2Te_3$ metasurface that supports a Mie resonance in the mid-infrared and a plasmonic resonance in the visible regime.[60] By tuning the nanostructure geometry, the device enables polarization-resolved and spectrally selective detection in a compact platform. Related dual-band photodetectors based on dielectric resonators embedded in metallic split-ring structures similarly demonstrate independently tunable plasmonic and Mie modes for multispectral infrared detection.[94] Intrinsic material duality can also provide spectral selectivity: crystalline silicon nanostructures, for example, support ultraviolet interband plasmons together with visible Mie resonances, offering complementary surface-sensitive and volume-confined responses.[95] A representative device example is shown in Fig. 5c, where Ho et al. developed hybrid Si-Al nanostructures for submicron-scale digital imaging pixels.[58,96] By combining hybrid plasmonic-Mie absorption with a Schottky junction at the Si-Al interface, the platform achieves wavelength-selective absorption together with efficient charge separation and photocarrier collection. Beyond replacing conventional dye-based color filters, this architecture brings additional capabilities to ultrahigh-density imaging, including polarization-sensitive detection, directional light reception, and ultraviolet selectivity.

In contrast to spectrally selective photodetection, which focuses primarily on separating optical response in wavelength space, geometry-driven absorption engineering is concerned with how the structure itself defines the interaction between light and matter. Through careful control of size, shape, and coupling geometry, hybrid nanostructures can reshape the excitation and hybridization of electric, magnetic, and higher-order multipolar modes, leading to stronger field localization and enhanced absorption. As illustrated in Fig. 5d, one representative example consists of Si nanodisks embedded in a perforated Au film, forming a metal–dielectric–metal cavity.[59] The use of geometrical parameters allows the device to achieve near unity absorptance over a wide range of spectrum, with absorptance levels not falling below 99.7%. Also, there is a slight shift towards red due to changes in the refractive index of the medium from 1.33 to 1.41. Hybrid nanoantennas formed by vertical integration, shown in Fig. 5e, provide another option. These systems are made up of metal and dielectrics in a vertical stack format, thus providing multiple resonances with different spatial distributions of electric fields.[97]

More generally, geometry-driven hybridization has also been extended to plasmonic-dielectric systems supporting bound states in the continuum, where strong confinement, high Q factors, and reduced mode volumes provide additional opportunities for sensing, nonlinear absorption, and controllable light-matter interaction.[98,99]

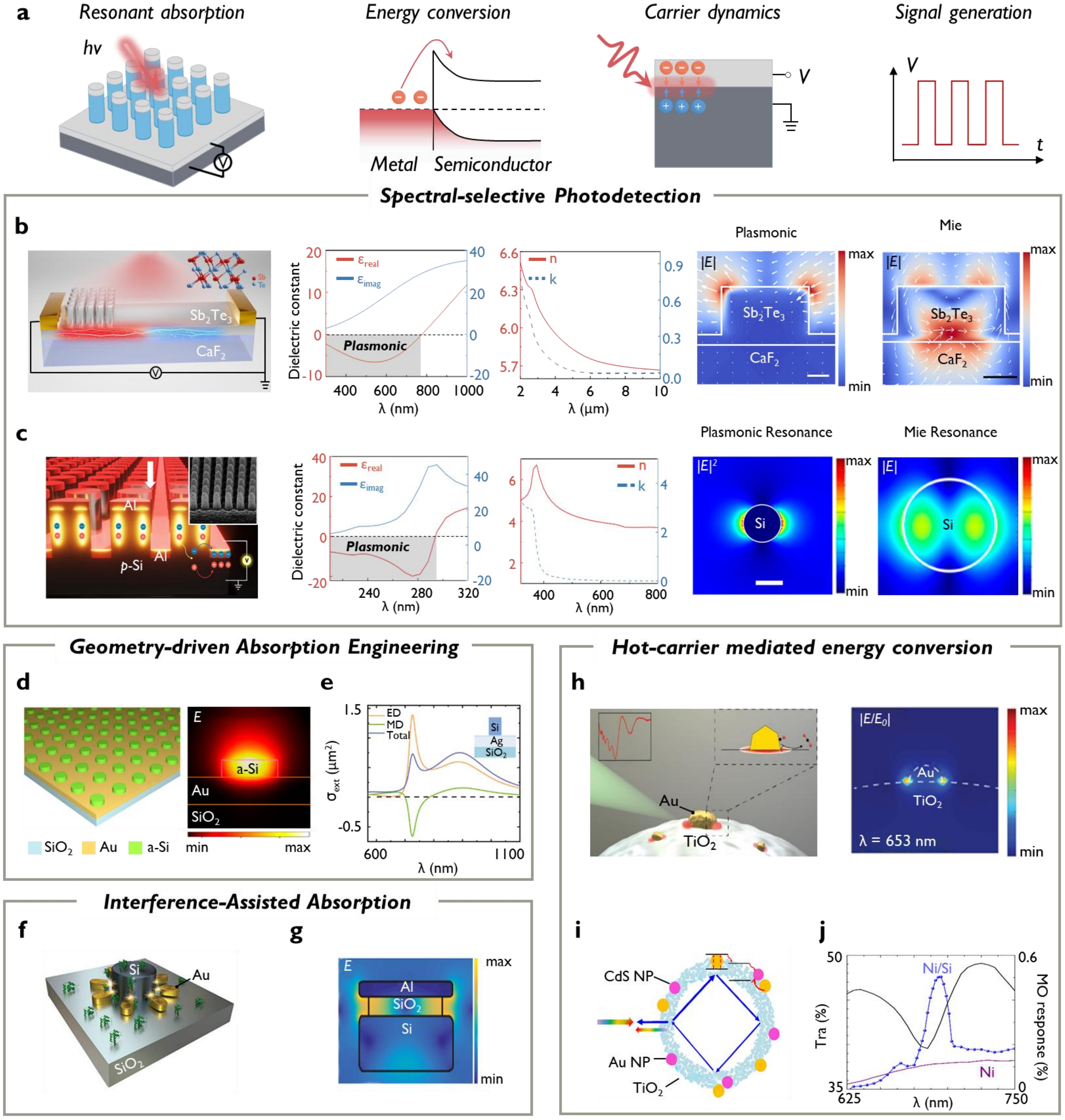

**Figure 5. Energy conversion and carrier dynamics in hybrid nanophotonic system. a.** Resonantly absorbed light is converted into electronic excitation, which drives subsequent carrier generation, transport, and relaxation, ultimately producing measurable output signals. **b.** $Sb_2Te_3$ metasurface photodetector illustrating plasmonic resonances in the visible regime and Mie resonances in the mid-infrared, with simulated field distributions and dielectric function of the $Sb_2Te_3$ film and the refractive index in the mid infrared.[60] **c**. Schematic of color-sensitive optical detectors based on the hybrid *p*-Si and Al nanostructures (left) and the spatial distribution of the absorption. Si disk with a diameter D = 120 nm. And in the middle the Silicon plasmon-Mie duality

showing plasmonic behavior in the UV and Mie resonance in the visible. [5,58,95] **d.** Schematic of the hybrid metasurface-based perfect absorber with a-Si on top of Au and $SiO_2$ layer and electric field distributions.[59] **e.** Hybrid gold-silicon nanoantenna's schematic. [143] **f.** A dimer of dual nanoresonators illuminated by circularly polarized light provides strong resonances.[100] **g.** Geometry of the hybrid metal-dielectric nanoantenna and the surface plots of electric field amplitude enhancement with 50 nm undercut.[101] **h**. Schematic of Mie resonant enhanced photocatalysis and electric field plots of bare $TiO_2$ and $TiO_2$ with 5 nm gold nanoparticle at 653 nm. [71] **i**. The schematic illustration of Mie resonance, light utilization, and charge separation in hollow $TiO_2$-Au-CdS ternary composites. [72] **j.** Experimental transmission (black, solid) and magneto-optical response (blue, circles) of the enhancement of magneto-optical effects in a nanoscale structure.[73]

Beyond structural tuning alone, interference-assisted absorption engineering exploits modal interference between electric, magnetic, and higher-order multipoles to shape absorption and scattering spectra. In the chiral-sensing platform shown in Fig. 5f, Mohammadi et al. used a dual-resonator hybrid plasmonic-Mie system in which near-field interference between spectrally overlapped electric and magnetic resonances with a $\pi/2$ phase difference enhances optical chirality and circular dichroism, enabling polarization-resolved detection.[100] A related strategy is illustrated in Fig. 5g, where Ray et al. designed Al-Si hybrid antennas in which dipole-quadrupole interference produces sharp spectral features with high refractive-index sensitivity.[101] More broadly, hybrid metasurfaces based on coupled plasmonic and dielectric building blocks further show that multipolar interference can be exploited to tailor scattering, absorption, and light redistribution in a highly controllable manner.[102]

When optical energy is strongly concentrated in metallic regions, resonance decay can generate nonequilibrium carriers that directly participate in energy transfer and conversion processes. This provides the basis for hot-carrier-mediated energy conversion, in which hybrid nanophotonic structures function not only as absorbers, but also as active platforms for harvesting and utilizing energetic charge carriers. In the hybrid $TiO_2$-Au-CdS photocatalyst shown in Fig. 5h, Hershey et al. demonstrated that the scattering response and associated photocatalytic behavior can be tuned by varying the inner diameter of the $TiO_2$ nanoshell.[71] In a related Au / TiO2 system (Fig. 5i), Yao et al. showed that Mie-resonance-assisted light confinement enhances CO oxidation photocatalysis by strengthening near-field hotspots and improving photon harvesting below 500 nm.[72] Hybrid energy-conversion concepts have also been extended beyond photocatalysis. In the magneto-optical platform shown in Fig. 5j, Barsukova et al. exploited magnetic Mie modes in a hybrid magnetophotonic stack to achieve a several-fold enhancement of nonreciprocal effects, establishing an externally tunable route toward active optical devices.[73] Together, these examples highlight how hot-carrier generation and resonance-assisted energy transfer can broaden the functional scope of hybrid nanophotonic systems well beyond conventional photodetection.

Overall, energy conversion and carrier dynamics in hybrid nanophotonic systems constitute a broad and versatile framework in which resonant absorption is directly linked to functional electronic, or optical outputs. Across spectral-selective photodetection, geometry-driven absorption engineering, interference-assisted absorption control, and hot-carrier-mediated conversion, hybrid plasmonic-Mie architectures offer multiple routes for regulating how optical energy is absorbed, transformed, and utilized at the nanoscale.

### 4.3. Radiative and Directional Emission Engineering

Beyond photodetection, hybrid plasmonic-Mie nanostructures also provide a versatile platform for controlling light emission. The same hybrid resonance principles that enhance light absorption can be harnessed to regulate radiative dynamics by tailoring the coupling between emitters and optical modes. High-index dielectric resonators support multipolar Mie modes and strong near fields, which, when combined with plasmonic elements or quantum emitters, enable enhanced spontaneous emission, active modulation, and directional radiation control.[28] As illustrated in Fig. 6a, incident excitation interacts with the hybrid resonator, which modifies the local electromagnetic environment and redistributes radiative channels, ultimately leading to engineered emission. On this basis, this section discusses three representative manifestations of radiative engineering in hybrid nanophotonic systems: Purcell enhancement (Fig. 6b,c), active emission modulation (Fig. 6d), and directional emission engineering (Fig. 6e-g).[103,104]

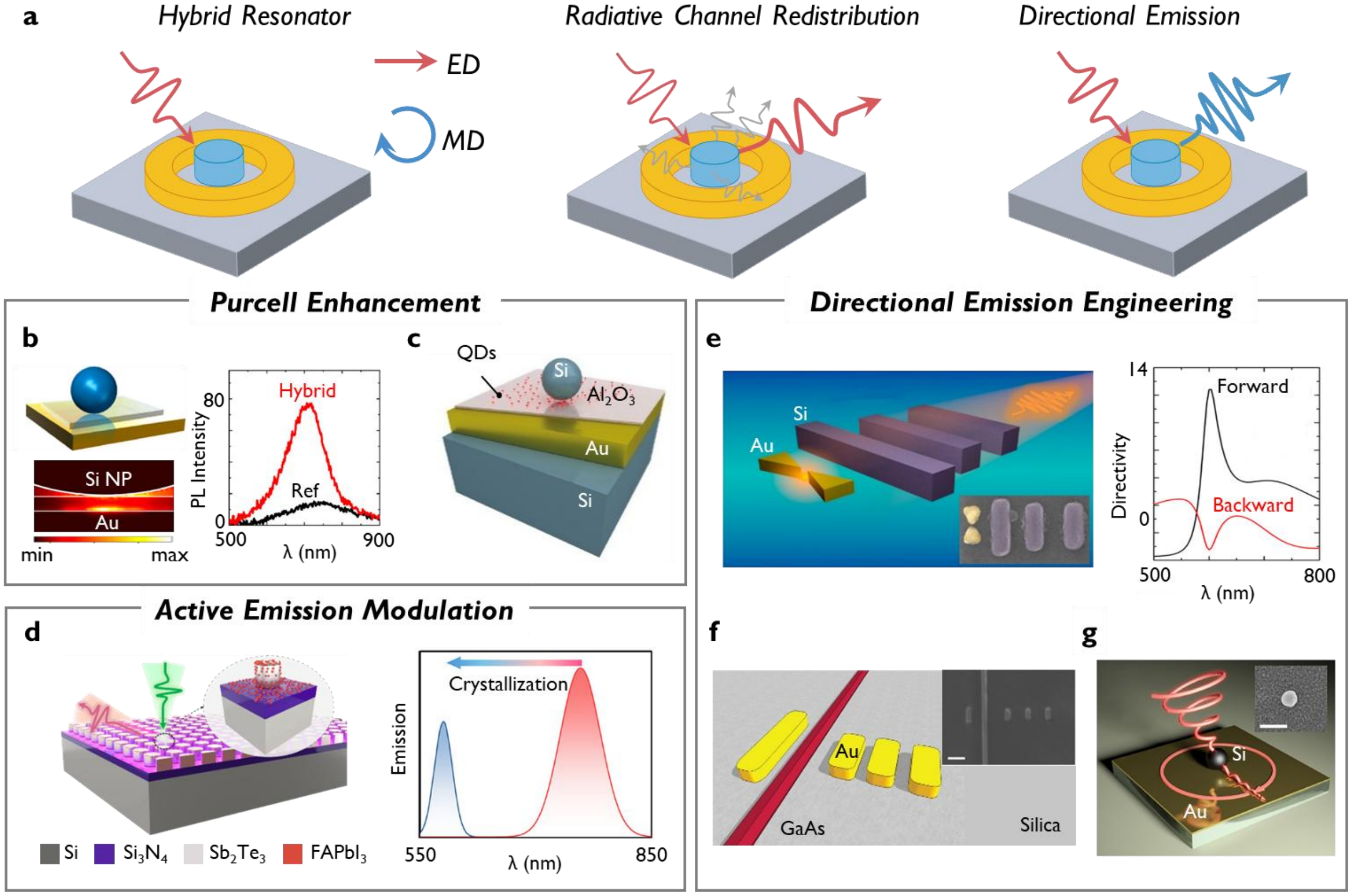

**Figure 6. Control of Spontaneous Emission and Radiation Patterns. a**. Schematic illustration of radiative and directional emission. Purcell Enhancement: **b.** Schematic of silicon nanoparticle-on-mirror (SiNPoM) and the photoluminescence spectra of a colloidal Si quantum dots monolayer on a Au film without (black curve) and with a Si nanoparticle (red curve) on it. [63] **c**. Schematic of the SiNPoM consisting of a crystalline spherical silicon nanoparticle (SiNP) situated on a gold film covered successively by a thin $Al_2O_3$ separation layer and a self-assembled CdSe/ZnS quantum dot monolayer[64] **c**. The design of the active silicon nanoparticles on metal.[105] Active Emission Modulation: **d**. Schematic illustration and operating principle of disk-based $Sb_2Te_3$ / perovskite hybrid nanoantennas. Perovskite quantum dots conformally coat $Sb_2Te_3$ nanodisks, enabling photoemission tuning via surface-enhanced Landau damping and hot-electron injection. Photoluminescence spectra demonstrate passive emission wavelength tuning, modulated by either the $Sb_2Te_3$ phase transition or an external electrical bias.[67] **e.** Schematic of the hybrid Au-Si Yagi-Uda nanoantenna with the calculated directivity of the optimized hybrid Yagi-Uda nanoantenna.[65] **f.** Schematic representation of a hybrid nanowire-plasmonic Yagi-Uda antenna and one of the designed Yagi-Uda nanoantenna radiation pattern.[66] **g.** Demonstration of the concept of plasmonic beam steering by silicon nanoantenna on top of gold substrate.

Purcell enhancement is a key manifestation of radiative engineering in hybrid nanophotonic systems, arising from the strong coupling between emitters and hybrid optical modes. In silicon nanoparticle-on-mirror architectures, coupling between the electric dipole of the dielectric nanoparticle and its image dipole in the metallic substrate supports gap-insensitive hybrid plasmonic-Mie modes, enabling broadband emission enhancement while mitigating the tunnelling-related losses associated with fully metallic counterparts, as shown in Fig. 6b.[63] Similar silicon nanoparticle platforms integrated with different quantum-dot layers likewise provide substantial spontaneous-emission enhancement through tailored Mie resonances,[64] while more complex hybrid architectures exploit multipolar and multimode interactions to further shape emission dynamics and spectral response.[104] In particle-on-metal geometries, the enhanced emission can also be funneled into surface plasmon polaritons, where electric-magnetic dipole interference enables directional launching and routing of guided modes.[105,106] [107,108,109] Together, these studies illustrate how hybrid plasmonic-Mie platforms provide a versatile route for enhancing and controlling spontaneous emission across a broad range of emitter-resonator systems.

Active emission modulation in hybrid nanophotonic systems extends radiative engineering beyond passive enhancement by enabling dynamic control over the emission wavelength and intensity. A representative example is provided by a disk-based hybrid metasurface in which perovskite quantum emitters are integrated with phase-change $Sb_2Te_3$ nanodisks, as shown in Fig. 6d. [67] Under optical excitation, localized interband plasmons in crystalline $Sb_2Te_3$ undergo surface-enhanced Landau damping.[110] These carriers can be efficiently injected into adjacent quantum emitters, thereby modifying carrier recombination pathways within the perovskite and producing a pronounced emission-energy shift exceeding 360 meV. Moreover, the emission response can be further tuned by an external bias, allowing active and reversible control over both the spectral position and intensity of the emitted light. This example highlights how hybrid plasmonic-dielectric platforms can combine resonant field enhancement with carrier-mediated interactions to realize actively tunable nanoscale light sources.

Directional emission engineering further extends hybrid radiative control to the shaping of angular radiation patterns, polarization characteristics, and optical routing. A particularly effective strategy is offered by Yagi-Uda-type hybrid nanoantennas, in which dielectric or semiconductor emitters are integrated with metallic directors and reflectors, as shown in Fig. 6e,f. [65,66] Beyond the Yagi-Uda-type hybrid nanoantennas, as shown in Fig. 6g, multi-resonant hybrid nanoantennas can combine electric, magnetic and higher-order multipolar responses to tailor far-field emission at the subwavelength scale.[106, 111, 112] In addition to the directional routing, hybrid emission platforms also support additional radiative functionalities, including enhanced nonlinear emission in nanowire-dimer systems and emission control mediated by high-Q supercavity and anapole-related hybrid modes in van der Waals nanoantennas.[113, 114]

In summary, hybrid plasmonic-Mie nanophotonic systems provide a powerful framework for radiative and directional emission engineering by combining the modal selectivity of dielectric Mie resonances with the strong field confinement of plasmonic structures. Through this synergy, these platforms enable not only enhanced spontaneous emission, but also active control over emission spectrum and precise shaping of radiation pathways.

### 4.4. Emerging Topological Material Platforms for Hybrid Plasmonic-Mie Resonances

Topological insulators (TIs), such as crystalline $Sb_2Te_3$, $Bi_2Te_3$, and $Bi_2Se_3$, are emerging as an important material platform for extending hybrid plasmonic–Mie physics beyond conventional metals and dielectrics.[115] Their defining feature is a dual electronic structure composed of an insulating bulk and metallic surface states protected by time-reversal symmetry. These surface states host massless Dirac fermions and are characterized by spin–momentum locking together with a single Dirac cone in the band structure.[115,116]

Because of this duality, TIs can simultaneously support dielectric Mie resonances in the high-index, low-loss bulk and plasmonic modes originating from the metallic surface states, while also allowing controlled coupling between them. Moreover, hot electrons generated under optical or electrical excitation provide an additional pathway for interfacial energy transfer and dynamic modulation.[67]

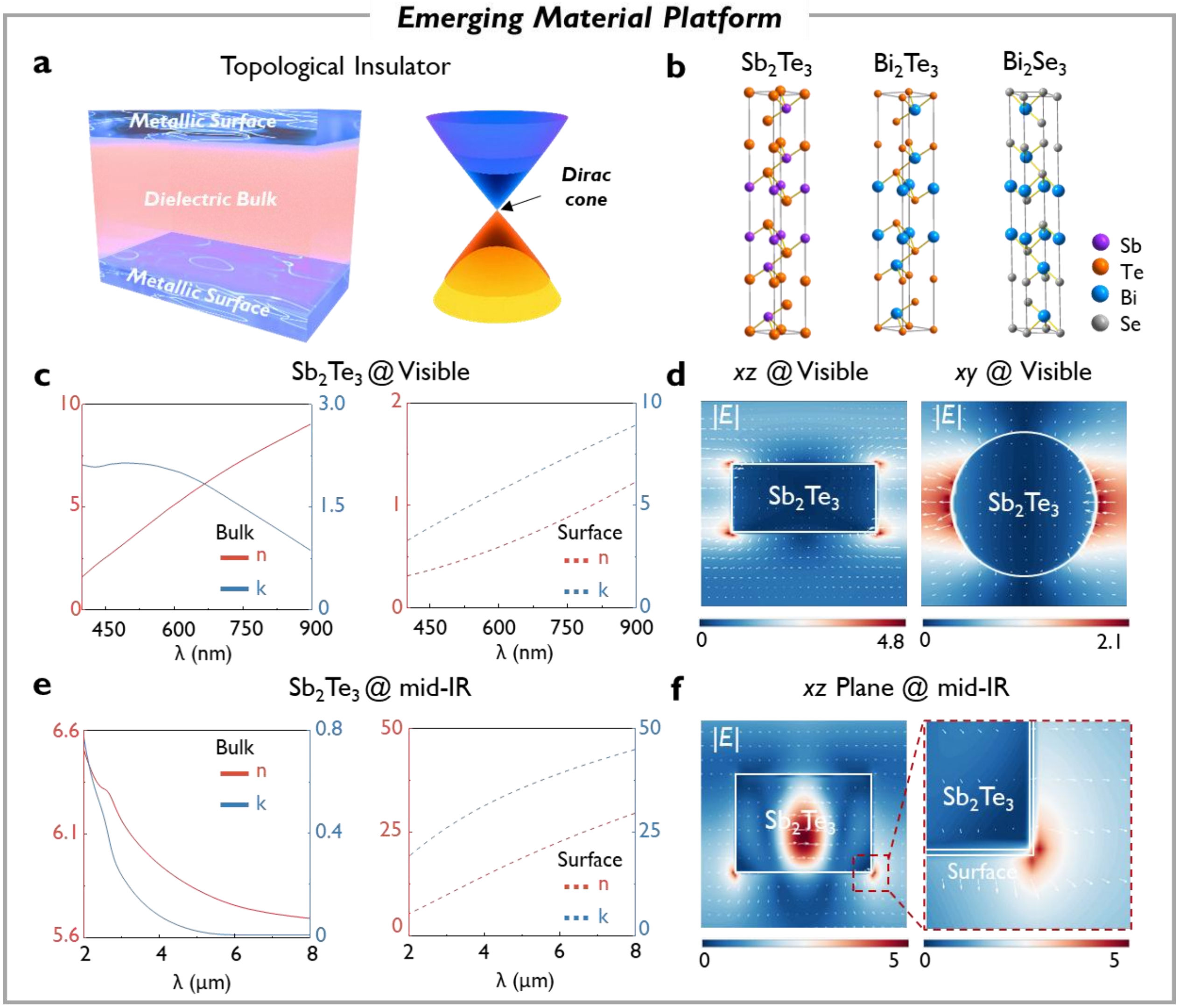

**Figure 7. Topological-insulator-based hybrid plasmonic–Mie platforms. a,** Schematic illustration of the key characteristics of topological insulators, including an insulating bulk, metallic surface states, and Dirac-cone dispersion. **b,** Atomic structures of $Sb_2Te_3$, $Bi_2Te_3$, and $Bi_2Se_3$, all built from quintuple-layer units. **c,e,** Optical constants ($n$ and $k$) of $Sb_2Te_3$ in different spectral regimes, showing the visible response (**c**), the experimentally measured mid-infrared bulk response (**e**, left), and the fitted mid-infrared surface response derived from experimental data (**e**, right). **d,** Simulated electric-field-intensity ($|E|$) distributions for a visible $Sb_2Te_3$ metasurface in the $xz$and $xy$planes, highlighting strong plasmonic resonance and surface-associated field localization. **f,** Simulated $|E|$ distributions for a mid-infrared $Sb_2Te_3$ metasurface in the $xz$plane (**left**) and at the enlarged interface region (**right**), showing simultaneous surface localization and volumetric confinement.

Fig. 7 brings together the key structural and optical features that support hybrid plasmonic–Mie resonances in topological-insulator (TI) materials. As shown in Fig. 7a, the fundamental physical properties of topological insulators are jointly determined by the bulk insulating state, the metallic surface state, and the corresponding Dirac cone band structure. The layered crystal structures of $Sb_2Te_3$, $Bi_2Te_3$, and $Bi_2Se_3$, as shown in Fig. 7b, provide an important structural foundation for the effective decoupling of surface states. Experimental measurements of the optical constants of $Sb_2Te_3$ (Fig. 7c and 7e) further reveal its significant frequency-dependent response characteristics. In the visible light region, its optical behavior exhibits strong metallic properties; however, as the wavelength extends into the mid-infrared region, its response gradually transitions to behavior dominated by dielectric properties, a trend that is particularly pronounced in the mid-infrared band. Numerical simulation results further validate this dual optical behavior. Figure 7d shows the electric field intensity ($|E|$) distribution of the $Sb_2Te_3$ metasurface in the visible light region. A significant localized field enhancement induced by the surface metallic state can be clearly observed, demonstrating typical plasmon resonance characteristics. In contrast, the mid-infrared $Sb_2Te_3$ metasurface shown in Fig. 7f exhibits a field distribution characterized by the coexistence of surface local enhancement and bulk mode confinement, indicating a hybrid response resulting from the mutual coupling of plasmon resonance and Mie resonance in this system. It is precisely this intrinsic coupling property of the material that enables topological insulators to possess multiple functions, including field enhancement, absorption control, and active modulation via surface carriers, thereby establishing their unique advantage as a material platform for reconfigurable hybrid photonic systems.

The representative experimental results are summarized in Tab. 2. Direct hybridization has been demonstrated in $Sb_2Te_3$ metasurfaces[60] and further confirmed by full-wave simulations in this work, showing coupled surface and bulk resonances with broadband field confinement. In contrast, $Bi_2Se_3$ nanobeams[116] exhibit high-index Mie modes with tunable near-field phases influenced by surface states, and $Bi_2Te_3$ nanoplates[117] support localized and dark plasmonic modes originating from topological surface currents. At the interfacial level, $Bi_2Se_3$-Au hybrids[118] and $Sb_2Te_3$-Perovskite antennas[67] reveal exciton / charge-mediated plasmon coupling and hot-electron transfer, enabling polarization-tunable or electrically controlled emission. Furthermore, $Bi_2Se_3$ / $Bi_2Te_3$ thin films or nanospheres[119] exhibit spin-polarized plasmonic modes leading to chiral terahertz emission and strong Purcell enhancement, establishing a link to spintronic photonics.

Together, these systems mark a clear transition from plasmonic- or Mie-dominant resonances to directly hybridized modes in topological metasurfaces. With their simultaneous access to strong field confinement, tunable spectral response, and spin-active electronic properties, topological insulators provide a promising foundation for reconfigurable, broadband, and quantum-compatible nanophotonic devices.

**Table 2. Representative topological-insulator systems associated with plasmonic and Mie resonances**

| Category | Platform | Mechanism | Functionality |
|---|---|---|---|
| Direct hybridization | $Sb_2Te_3$ [60] | Plasmon-dielectric hybridization | Broadband photodetection |
| | $Sb_2Te_3$ (simulation) | Surface localization + volumetric confinement | Concept validation |
| Indirect | $Bi_2Se_3$ [116] | High-index Mie modes 2π near-field phase | Phase / nonlinear control |
| | $Bi_2Te_3$ [117,120] | Topological surface plasmons | Structural coloration |
| Interfacial | $Bi_2Se_3$-Au [118] | Exciton/charge-plasmon coupling | Polarization-tunable emission |
| | $Sb_2Te_3$-Perovskite[67] | Hot-electron transfer, emission tuning | Quantum light source |
| | $Bi_2Se_3$ / $Bi_2Te_3$[119] | Spin-polarized plasmons, chiral THz | Spintronic photonics |

## 5. Conclusion and outlook

Hybrid plasmonic–Mie resonance systems are emerging as a key frontier in the field of nanophotonics. By ingeniously combining the strong near-field confinement capabilities of plasmons with the low-loss and multi-degree-of-freedom tunability of Mie resonance, these systems have successfully overcome the inherent limitations of traditional all-metal or all-dielectric resonators in terms of loss, field confinement, and functional expansion. This complementary interaction and coupling not only enables unprecedented depth of light field manipulation at the nanoscale but also significantly broadens the scope of fundamental research into novel light–matter interactions. Currently, this research direction has demonstrated immense application potential in fields such as ultra-sensitive sensing, reconfigurable metasurfaces, and on-chip integrated photonic devices.

Moreover, the application of the near-zero permittivity (ENZ) structure adds an entirely new physical dimension to the realm of hybrid resonance phenomena. The high dispersion property of ENZ materials close to the zero-permittivity threshold allows these materials to offer significant benefits in terms of ultra-localized confinement of the electromagnetic field, exact phase management, and reshaping of the energy flow. This approach not only extends the physical significance of hybrid resonances but also opens new directions of research towards understanding novel mode couplings, nonlinear optical effects, and unusual concepts in light manipulation.

Looking ahead, advancing hybrid resonance devices from proof-of-concept to high-performance, scalable on-chip integrated applications will rely heavily on the synergistic evolution of nanofabrication technologies, inverse design algorithms, and novel quantum/topological materials. Among numerous emerging platforms, topological insulators have transcended the scope of traditional metal-dielectric systems due to their intrinsic property of coexisting "insulator bulk states and metallic surface states." Combining multiple advantages such as plasmonic field enhancement, high-refractive-index dielectric confinement, and topologically protected transport, they hold promise for fostering entirely new hybrid resonance modes that integrate spin selectivity, low loss, and dynamic tunability. Simultaneously, the ENZ platform will play an indispensable role in mode engineering, broadband phase modulation, and energy flow reorganization within metal-free or weakly metallized hybrid systems.

In summary, hybrid plasmonic–Mie resonance systems have established a physical framework for nanoscale light field control that combines flexibility with universality. With the continuous evolution of ENZ materials, topological insulators, active control mechanisms, and photonic integration architectures, this field will provide a core driving force for the development of quantum photonics, spin-optics, and broadband optoelectronic devices, and will have a profound impact on cutting-edge applications such as next-generation high-performance sensing, high-resolution imaging, optical communications, and all-optical information processing.

**Acknowledgments**

Z.D. and J.K.W.Y. would like to acknowledge the funding support from The National Research Foundation (NRF), Singapore via Grant No. NRF-CRP30-2023-0003. In addition, Z.D. would like to acknowledge the funding support from the Agency for Science, Technology and Research (A*STAR) under MTC IRG (Project No. M24N7c0083), and the SUTD Kickstarter Initiative (SKI 2021-06-05) SUTD Kickoff. J.K.W.Y. would like to acknowledge the funding support from National Research Funding (NRF) Singapore NRF-CRP20-2017-0001 and NRF-NRFI06-2020-0005. This research is also supported by grants from the National Research Foundation, Prime Minister's Office, Singapore under its Campus of Research Excellence and Technological Enterprise (CREATE) programme. Y. L. would like to acknowledge the funding support from the Agency for Science, Technology and Research (A*STAR) under the Career Development Award grant (Project No. C243512003). J.S. would like to acknowledge the funding support from the National Research Foundation of Korea (NRF) (grant no. RS-2024-00356928) funded by the Ministry of Science and ICT (MSIT) of the Korean government. J.S. acknowledges the Hyundai Motor Chung Mong-Koo Foundation; the Institute of Information & Communications Technology Planning & Evaluation (IITP) (grant no. RS-2019-II191906; POSTECH Artificial Intelligence Graduate School Program) funded by the MSIT of the Korean government; and the NRF Ph.D. fellowship (grant no. RS-2025-25438817) funded by the Ministry of Education (MOE) of the Korean government.

**Authors' contributions**

Z. D., J. K. W. Y., and Y. K. conceived the review concept, organized the paper sections, and supervised the project. S. Z. and C.-F. P. wrote the manuscript and created the figures. Y. F. was responsible for on-chip hybrid resonators domain. J. S. and J. R. were responsible for the aspect of Machine Learning and Data-Driven Approaches. Yuanda. L. and J. W. were responsible for the aspect of Topological Material Platforms. Yan L. contributed to the aspect of Photoemission. Y. L., J. D., J. R. and Y. K. participated in the discussions and provided suggestions. S. Z. and C.-F. P. contributed equally to this work. All authors have read and approved the final manuscript.

**Conflict of interest:** Authors state no conflicts of interest.

**Data availability:** Data sharing is not applicable to this article as no datasets were generated or analyzed during the current study.

**References**

1 Rezaei, S. D. *et al.* Tri-functional metasurface enhanced with a physically unclonable function. *Materials Today* **62**, 51-61 (2023).

2 Ha, S. T., Li, Q., Yang, J. K., Demir, H. V., Brongersma, M. L. & Kuznetsov, A. I. Optoelectronic metadevices. *Science* **386**, eadm7442 (2024).

3 Prasad, P. N. *Nanophotonics*. (John Wiley & Sons, 2004).

4 Chengfeng, P., Shutao, Z., Farsari, M., Oh, S. H. & Yang, J. K. Vol. 12 1359-1361 (De Gruyter, 2023).

5 Dong, Z. *et al.* Printing beyond sRGB color gamut by mimicking silicon nanostructures in free-space. *Nano letters* **17**, 7620-7628 (2017).

6 Rezaei, S. D. *et al.* Wide-gamut plasmonic color palettes with constant subwavelength resolution. *ACS nano* **13**, 3580-3588 (2019).

7 Liu, Y. *et al.* Structural color three-dimensional printing by shrinking photonic crystals. *Nature communications* **10**, 4340 (2019).

8 Liu, H. *et al.* High-order photonic cavity modes enabled 3D structural colors. *ACS nano* **16**, 8244-8252 (2022).

9 Dong, Z. *et al.* Schrödinger's red pixel by quasi-bound-states-in-the-continuum. *Science Advances* **8**, eabm4512 (2022).

10 Rezaei, S. D. *et al.* Nanophotonic structural colors. *ACS photonics* **8**, 18-33 (2021).

11 Gu, J. *et al.* Structural colors based on diamond metasurface for information encryption. *Advanced Optical Materials* **11**, 2202826 (2023).

12 Intaravanne, Y. & Chen, X. Recent advances in optical metasurfaces for polarization detection and engineered polarization profiles. *Nanophotonics* **9**, 1003-1014 (2020).

13 Juliano Martins, R. *et al.* Metasurface-enhanced light detection and ranging technology. *Nature communications* **13**, 5724 (2022).

14 Li, J., Li, J., Zhou, S. & Yi, F. Metasurface photodetectors. *Micromachines* **12**, 1584 (2021).

15 Jung, C. *et al.* Metasurface-driven optically variable devices. *Chemical Reviews* **121**, 13013-13050 (2021).

16 Hu, J., Lawrence, M. & Dionne, J. A. High quality factor dielectric metasurfaces for ultraviolet circular dichroism spectroscopy. *ACS Photonics* **7**, 36-42 (2019).

17 Chen, Q., Hu, X., Wen, L., Yu, Y. & Cumming, D. R. Nanophotonic image sensors. *Small* **12**, 4922-4935 (2016).

18 Qin, J. *et al.* Metasurface micro/nano-optical sensors: principles and applications. *ACS nano* **16**, 11598-11618 (2022).

19 Altug, H., Oh, S.-H., Maier, S. A. & Homola, J. Advances and applications of nanophotonic biosensors. *Nature nanotechnology* **17**, 5-16 (2022).

20 Gao, L., Qu, Y., Wang, L. & Yu, Z. Computational spectrometers enabled by nanophotonics and deep learning. *Nanophotonics* **11**, 2507-2529 (2022).

21 Yang, Z., Albrow-Owen, T., Cai, W. & Hasan, T. Miniaturization of optical spectrometers. *Science* **371**, eabe0722 (2021).

22 Cueff, S., Berguiga, L. & Nguyen, H. S. Fourier imaging for nanophotonics. *Nanophotonics* **13**, 841-858 (2024).

23 Pan, C.-F. *et al.* 3D-printed multilayer structures for high–numerical aperture achromatic metalenses. *Science advances* **9**, eadj9262 (2023).

24 Hao, H. *et al.* Single-shot 3D imaging meta-microscope. *Nano Letters* **24**, 13364-13373 (2024).

25 Meng, Y. *et al.* Optical meta-waveguides for integrated photonics and beyond. *Light: Science & Applications* **10**, 235 (2021).

26 Liu, G. *et al.* Single-photon generation and manipulation in quantum nanophotonics. *Applied Physics Reviews* **12** (2025).

27 González-Tudela, A., Reiserer, A., García-Ripoll, J. J. & García-Vidal, F. J. Light–matter interactions in quantum nanophotonic devices. *Nature Reviews Physics* **6**, 166-179 (2024).

28 Cai, H. *et al.* Charge-depletion-enhanced WSe2 quantum emitters on gold nanogap arrays with near-unity quantum efficiency. *Nature Photonics* **18**, 842-847 (2024).

29 Dong, Z. *et al.* Silicon nanoantenna mix arrays for a trifecta of quantum emitter enhancements. *Nano Letters* **21**, 4853-4860 (2021).

30 Xu, J. *et al.* Multiphoton upconversion enhanced by deep subwavelength near-field confinement. *Nano Letters* **21**, 3044-3051 (2021).

31 Krasnok, A. E. *et al.* Optical nanoantennas. *Physics-Uspekhi* **56**, 539 (2013).

32 Zhang, L., Mei, S., Huang, K. & Qiu, C. W. Advances in full control of electromagnetic waves with metasurfaces. *Advanced Optical Materials* **4**, 818-833 (2016).

33 Luo, X. Subwavelength artificial structures: opening a new era for engineering optics. *Advanced Materials* **31**, 1804680 (2019).

34 Lalanne, P., Yan, W., Vynck, K., Sauvan, C. & Hugonin, J. P. Light interaction with photonic and plasmonic resonances. *Laser & Photonics Reviews* **12**, 1700113 (2018).

35 Chen, Y. & Ming, H. Review of surface plasmon resonance and localized surface plasmon resonance sensor. *Photonic Sensors* **2**, 37-49 (2012).

36 Hutter, E. & Fendler, J. H. Exploitation of localized surface plasmon resonance. *Advanced materials* **16**, 1685-1706 (2004).

37 Zhao, Q., Zhou, J., Zhang, F. & Lippens, D. Mie resonance-based dielectric metamaterials. *Materials today* **12**, 60-69 (2009).

38 Dorodnyy, A., Smajic, J. & Leuthold, J. Mie scattering for photonic devices. *Laser & Photonics Reviews* **17**, 2300055 (2023).

39 Mie, G. Beiträge zur Optik trüber Medien, speziell kolloidaler Metallösungen. *Annalen der physik* **330**, 377-445 (1908).

40 Barreda, Á., Vitale, F., Minovich, A. E., Ronning, C. & Staude, I. Applications of hybrid metal-dielectric nanostructures: state of the art. *Advanced Photonics Research* **3**, 2100286 (2022).

41 Guan, J., Park, J.-E., Deng, S., Tan, M. J., Hu, J. & Odom, T. W. Light–matter interactions in hybrid material metasurfaces. *Chemical Reviews* **122**, 15177-15203 (2022).

42 Lepeshov, S. I., Krasnok, A. E., Belov, P. A. & Miroshnichenko, A. E. Hybrid nanophotonics. *Physics-Uspekhi* **61**, 1035 (2019).

43 Bohren, C. F. & Huffman, D. R. *Absorption and scattering of light by small particles*. (John Wiley & Sons, 2008).

44 Johnson, P. B. & Christy, R.-W. Optical constants of the noble metals. *Physical review B* **6**, 4370 (1972).

45 Fu, Y. H., Kuznetsov, A. I., Miroshnichenko, A. E., Yu, Y. F. & Luk'yanchuk, B. Directional visible light scattering by silicon nanoparticles. *Nature communications* **4**, 1527 (2013).

46 Zhang, L. *et al.* Dynamically configurable hybridization of plasmon modes in nanoring dimer arrays. *Nanoscale* **7**, 12018-12022 (2015).

47 Fan, S., Suh, W. & Joannopoulos, J. D. Temporal coupled-mode theory for the Fano resonance in optical resonators. *Journal of the Optical Society of America A* **20**, 569-572 (2003).

48 Suh, W., Wang, Z. & Fan, S. Temporal coupled-mode theory and the presence of non-orthogonal modes in lossless multimode cavities. *IEEE Journal of Quantum Electronics* **40**, 1511-1518 (2004).

49 Yariv, A. Coupled-mode theory for guided-wave optics. *IEEE Journal of Quantum Electronics* **9**, 919-933 (1973).

50 Decker, M., Pertsch, T. & Staude, I. Strong coupling in hybrid metal–dielectric nanoresonators. *Philosophical Transactions of the Royal Society A: Mathematical, Physical and Engineering Sciences* **375** (2017).

51 Shi, T. *et al.* Planar chiral metasurfaces with maximal and tunable chiroptical response driven by bound states in the continuum. *Nature Communications* **13**, 4111 (2022).

52 Martens, K., Funck, T., Santiago, E. Y., Govorov, A. O., Burger, S. & Liedl, T. Onset of chirality in plasmonic meta-molecules and dielectric coupling. *ACS nano* **16**, 16143-16149 (2022).

53 Yue, W., Gao, S., Lee, S.-S., Kim, E.-S. & Choi, D.-Y. Subtractive color filters based on a silicon-aluminum hybrid-nanodisk metasurface enabling enhanced color purity. *Scientific reports* **6**, 29756 (2016).

54 Yue, W., Gao, S., Lee, S. S., Kim, E. S. & Choi, D. Y. Highly reflective subtractive color filters capitalizing on a silicon metasurface integrated with nanostructured aluminum mirrors. *Laser & Photonics Reviews* **11**, 1600285 (2017).

55 Zhang, S. *et al.* Reversible electrical switching of nanostructural color pixels. *Nanophotonics* **12**, 1387-1395 (2023).

56 Kim, Y. & Hyun, J. K. Encoding Mie, plasmonic, and diffractive structural colors in the same pixel. *Nanophotonics* **12**, 3341-3349 (2023).

57 Nishi, H. & Tatsuma, T. Full-color scattering based on plasmon and mie resonances of gold nanoparticles modulated by Fabry–Pérot interference for coloring and image projection. *ACS Applied Nano Materials* **2**, 5071-5078 (2019).

58 Ho, J. *et al.* mezing color-sensitive photodetectors via hybrid nanoantennas toward submicrometer dimensions. *Science Advances* **8**, eadd3868 (2022).

59 Aoni, R. A., Rahmani, M., Xu, L. & Miroshnichenko, A. Hybrid Metasurface Based Tunable Near-Perfect Absorber and Plasmonic Sensor. (2018).

60 Zhang, S. *et al.* Chalcogenide Metasurfaces Enabling Ultra-Wideband Detectors From Visible to Mid-infrared. *Advanced Science* **12**, 2413858 (2025).

61 Sun, X., Dai, D., Thylén, L. & Wosinski, L. in *Photonics.* 1116-1130 (MDPI).

62 Zhou, L., Sun, X., Li, X. & Chen, J. Miniature microring resonator sensor based on a hybrid plasmonic waveguide. *Sensors* **11**, 6856-6867 (2011).

63 Sugimoto, H. & Fujii, M. Broadband dielectric–metal hybrid nanoantenna: Silicon nanoparticle on a mirror. *Acs Photonics* **5**, 1986-1993 (2018).

64 Yang, G., Niu, Y., Wei, H., Bai, B. & Sun, H.-B. Greatly amplified spontaneous emission of colloidal quantum dots mediated by a dielectric-plasmonic hybrid nanoantenna. *Nanophotonics* **8**, 2313-2319 (2019).

65 Ho, J. *et al.* Highly directive hybrid metal–dielectric Yagi-Uda nanoantennas. *ACS nano* **12**, 8616-8624 (2018).

66 Ramezani, M. *et al.* Hybrid semiconductor nanowire–metallic Yagi-Uda antennas. *Nano letters* **15**, 4889-4895 (2015).

67 Liu, Y. *et al.* Electrically Tunable and Modulated Perovskite Quantum Emitters via Surface-Enhanced Landau Damping. *Advanced Materials* **37**, 2419076 (2025).

68 Yao, D., Zhang, Y., Huang, Y., Liu, Y., Han, G. & Hao, Y. Ultra high Q and widely tunable Mie resonance nanofilter at telecom wavelengths. *IEEE Photonics Journal* **13**, 1-9 (2021).

69 Wu, Y., Kang, L., Bao, H. & Werner, D. H. Exploiting topological properties of Mie-resonance-based hybrid metasurfaces for ultrafast switching of light polarization. *Acs Photonics* **7**, 2362-2373 (2020).

70 Kim, S. *et al.* Strong Light Confinement in Metal-Coated Si Nanopillars: Interplay of Plasmonic Effects and Geometric Resonance. *Nanoscale research letters* **12**, 1-5 (2017).

71 Hershey, M., Lu, G., North, J. D. & Swearer, D. F. Mie Resonant Metal Oxide Nanospheres for Broadband Photocatalytic Enhancements. *ACS nano* **18**, 18493-18502 (2024).

72 Yao, X. *et al.* Mie resonance in hollow nanoshells of ternary TiO2-Au-CdS and enhanced photocatalytic hydrogen evolution. *Applied Catalysis B: Environmental* **276**, 119153 (2020).

73 Barsukova, M. G., Shorokhov, A. S., Musorin, A. I., Neshev, D. N., Kivshar, Y. S. & Fedyanin, A. A. Magneto-optical response enhanced by Mie resonances in nanoantennas. *Acs Photonics* **4**, 2390-2395 (2017).

74 Chen, W. *et al.* Broadband solar metamaterial absorbers empowered by transformer-based deep learning. *Advanced Science* **10**, 2206718 (2023).

75 Chen, W. *et al.* All-Dielectric SERS Metasurface with Strong Coupling Quasi-BIC Energized by Transformer-Based Deep Learning. *Advanced Optical Materials* **12**, 2301697 (2024).

76 Gao, Y. *et al.* Meta-attention deep learning for smart development of metasurface sensors. *Advanced Science* **11**, 2405750 (2024).

77 Ma, W. *et al.* Deep learning empowering design for selective solar absorber. *Nanophotonics* **12**, 3589-3601 (2023).

78 Chen, Y. *et al.* Machine-learning-assisted photonic device development: a multiscale approach from theory to characterization. *Nanophotonics* (2025).

79 Han, J. H. *et al.* Neural-network-enabled design of a chiral plasmonic nanodimer for target-specific chirality sensing. *ACS nano* **17**, 2306-2317 (2023).

80 Ahmed, W. W. *et al.* Machine learning assisted plasmonic metascreen for enhanced broadband absorption in ultra-thin silicon films. *Light: Science & Applications* **14**, 42 (2025).

81 Decker, M. *et al.* Dual-channel spontaneous emission of quantum dots in magnetic metamaterials. *Nature communications* **4**, 2949 (2013).

82 Barati Sedeh, H. *et al.* Manipulation of scattering spectra with topology of light and matter. *Laser & Photonics Reviews* **17**, 2200472 (2023).

83 Yao, K., Unni, R. & Zheng, Y. Intelligent nanophotonics: merging photonics and artificial intelligence at the nanoscale. *Nanophotonics* **8**, 339-366 (2019).

84 Wiecha, P. R. & Muskens, O. L. Deep learning meets nanophotonics: a generalized accurate predictor for near fields and far fields of arbitrary 3D nanostructures. *Nano letters* **20**, 329-338 (2019).

85 Lynch, S. *et al.* Physics-Guided Hierarchical Neural Networks for Maxwell's Equations in Plasmonic Metamaterials. *ACS photonics* (2025).

86 Lee, D., Zhang, L., Yu, Y. & Chen, W. Deep neural operator enabled concurrent multitask design for multifunctional metamaterials under heterogeneous fields. *Advanced Optical Materials* **12**, 2303087 (2024).

87 Ma, W., Cheng, F. & Liu, Y. Deep-learning-enabled on-demand design of chiral metamaterials. *ACS nano* **12**, 6326-6334 (2018).

88 Zhang, Z., Yang, C., Qin, Y., Feng, H., Feng, J. & Li, H. Diffusion probabilistic model based accurate and high-degree-of-freedom metasurface inverse design. *Nanophotonics* **12**, 3871-3881 (2023).

89 Ma, W., Cheng, F., Xu, Y., Wen, Q. & Liu, Y. Probabilistic representation and inverse design of metamaterials based on a deep generative model with semi-supervised learning strategy. *Advanced Materials* **31**, 1901111 (2019).

90 So, S. & Rho, J. Designing nanophotonic structures using conditional deep convolutional generative adversarial networks. *Nanophotonics* **8**, 1255-1261 (2019).

91 Babicheva, V. E. & Evlyukhin, A. B. Mie-resonant metaphotonics. *Advances in Optics and Photonics* **16**, 539-658 (2024).

92 Nagasaki, Y., Hotta, I., Suzuki, M. & Takahara, J. Metal-masked mie-resonant full-color printing for achieving free-space resolution limit. *Acs Photonics* **5**, 3849-3855 (2018).

93 Holsteen, A. L., Raza, S., Fan, P., Kik, P. G. & Brongersma, M. L. Purcell effect for active tuning of light scattering from semiconductor optical antennas. *Science* **358**, 1407-1410 (2017).

94 Wang, C., Huang, M., Zhang, Z. & Xu, W. Dual band metamaterial absorber: Combination of plasmon and Mie resonances. *Journal of Materials Science & Technology* **53**, 37-40 (2020).

95 Dong, Z. *et al.* Ultraviolet interband plasmonics with Si nanostructures. *Nano Letters* **19**, 8040-8048 (2019).

96 Chen, W. *et al.* Enabling highly efficient infrared silicon photodetectors via disordered metasurfaces with upconversion nanoparticles. *Science Advances* **11**, eadx7783 (2025).

97 McPolin, C. P., Vila, Y. N., Krasavin, A. V., Llorca, J. & Zayats, A. V. Multimode hybrid gold-silicon nanoantennas for tailored nanoscale optical confinement. *Nanophotonics* **12**, 2997-3005 (2023).

98 Xiang, J., Xu, Y., Chen, J.-D. & Lan, S. Tailoring the spatial localization of bound state in the continuum in plasmonic-dielectric hybrid system. *Nanophotonics* **9**, 133-142 (2020).

99 Soliman, A., Williams, C., Hopper, R., Udrea, F., Butt, H. & Wilkinson, T. D. High-Transmission Mid-Infrared Bandpass Filters using Hybrid Metal-Dielectric Metasurfaces for CO2 Sensing. *Advanced Optical Materials* **13**, 2402603 (2025).

100 Mohammadi, E., Tittl, A., Tsakmakidis, K. L., Raziman, T. & Curto, A. G. Dual nanoresonators for ultrasensitive chiral detection. *ACS photonics* **8**, 1754-1762 (2021).

101 Ray, D., Raziman, T., Santschi, C., Etezadi, D., Altug, H. & Martin, O. J. Hybrid metal-dielectric metasurfaces for refractive index sensing. *Nano Letters* **20**, 8752-8759 (2020).

102 Guo, R. *et al.* Multipolar coupling in hybrid metal–dielectric metasurfaces. *Acs Photonics* **3**, 349-353 (2016).

103 Bidault, S., Mivelle, M. & Bonod, N. Dielectric nanoantennas to manipulate solid-state light emission. *Journal of Applied Physics* **126** (2019).

104 Liu, T., Xu, R., Yu, P., Wang, Z. & Takahara, J. Multipole and multimode engineering in Mie resonance-based metastructures. *Nanophotonics* **9**, 1115-1137 (2020).

105 Yaroshenko, V., Zuev, D. & Evlyukhin, A. B. Resonant channeling of light near metal surface by passive and active silicon nanoparticles. *Surfaces and Interfaces* **34**, 102344 (2022).

106 Sinev, I., Komissarenko, F., Iorsh, I., Permyakov, D., Samusev, A. & Bogdanov, A. Steering of guided light with dielectric nanoantennas. *ACS Photonics* **7**, 680-686 (2020).

107 Sinev, I. S. *et al.* Chirality driven by magnetic dipole response for demultiplexing of surface waves. *Laser & Photonics Reviews* **11**, 1700168 (2017).

108 Evlyukhin, A. B. & Bozhevolnyi, S. I. Resonant unidirectional and elastic scattering of surface plasmon polaritons by high refractive index dielectric nanoparticles. *Physical Review B* **92**, 245419 (2015).

109 Sun, S., Li, M., Du, Q., Png, C. E. & Bai, P. Metal–dielectric hybrid dimer nanoantenna: Coupling between surface plasmons and dielectric resonances for fluorescence enhancement. *The Journal of Physical Chemistry C* **121**, 12871-12884 (2017).

110 Clavero, C. Plasmon-induced hot-electron generation at nanoparticle/metal-oxide interfaces for photovoltaic and photocatalytic devices. *Nature Photonics* **8**, 95-103 (2014).

111 Dmitriev, P. A. *et al.* Hybrid dielectric-plasmonic nanoantenna with multiresonances for subwavelength photon sources. *ACS Photonics* **10**, 582-594 (2023).

112 Barreda, A., Hell, S., Weissflog, M., Minovich, A., Pertsch, T. & Staude, I. Metal, dielectric and hybrid nanoantennas for enhancing the emission of single quantum dots: A comparative study. *Journal of Quantitative Spectroscopy and Radiative Transfer* **276**, 107900 (2021).

113 Casadei, A. *et al.* Photonic–plasmonic coupling of GaAs single nanowires to optical nanoantennas. *Nano letters* **14**, 2271-2278 (2014).

114 Randerson, S. A. *et al.* High Q hybrid Mie–plasmonic resonances in van der Waals nanoantennas on gold substrate. *ACS nano* **18**, 16208-16221 (2024).

115 Zhang, H., Liu, C.-X., Qi, X.-L., Dai, X., Fang, Z. & Zhang, S.-C. Topological insulators in Bi2Se3, Bi2Te3 and Sb2Te3 with a single Dirac cone on the surface. *Nature physics* **5**, 438-442 (2009).

116 Nandi, S. *et al.* Unveiling Local Optical Properties Using Nanoimaging Phase Mapping in High-Index Topological Insulator Bi2Se3 Resonant Nanostructures. *Nano Letters* **23**, 11501-11509 (2023).

117 Zhao, M. *et al.* Visible surface plasmon modes in single Bi2Te3 nanoplate. *Nano letters* **15**, 8331-8335 (2015).

118 Bhattacharya, T. S., Raha, S., Mondal, P. K., Pradhan, M., Ghosh, S. & Singha, A. Tuning Light-Matter Interaction in 2D Bi2Se3 Through Plasmonic Particle Coupling. *Advanced Functional Materials*, e10990 (2025).

119 Zhao, Z., Wang, H., Hu, G. & Alù, A. Topological and Reconfigurable Terahertz Metadevices. *Research* **8**, 0882 (2025).

120 Krishnamoorthy, H., Adamo, G., Yin, J., Savinov, V., Zheludev, N. & Soci, C. Infrared dielectric metamaterials from high refractive index chalcogenides. *Nature Communications* **11**, 1692 (2020).